\documentclass[a4paper,11pt]{article}
\pdfoutput=1 

\usepackage{jcappub} 

\usepackage[T1]{fontenc} 

\title{\boldmath Dark Matter from Holography}

\author[]{Oem Trivedi,}
\author[]{Robert J. Scherrer}

\affiliation[]{Department of Physics and Astronomy, Vanderbilt University, Nashville, TN 37235, USA}
\emailAdd{oem.trivedi@vanderbilt.edu, robert.scherrer@vanderbilt.edu}

\abstract{Previous studies have examined the holographic principle as a means of producing dark energy.  Here we propose instead the possibility of holographic dark matter.  In this case,
dark matter does not arise in the framework of particle physics but is derived from the infrared cutoff set by the horizon scale.  Using the Ricci cutoff, and a universe containing only baryons and radiation, we can account for the dark matter and naturally explain the coincidence between baryonic and nonbaryonic contributions to the density.  In the presence of a pre-existing vacuum energy density our model reverses the sign of this density, thus accounting for the fact that certain string theories generically predict a negative vacuum energy, but observations require a positive value.}

\begin{document}
\maketitle
\flushbottom

\section{Introduction}

Observations over the past few decades have given a standard model of cosmology, in which the universe has a flat geometry with $\Omega = 1$ and consists of roughly 5\% baryonic matter, 25\% nonbaryonic dark matter \cite{dm11rubin1970rotation,dm1Cirelli:2024ssz,dm2Arbey:2021gdg}, and 70\% dark energy \cite{SupernovaSearchTeam:1998fmf,Perlmutter:1998np}. This cosmological model which is often referred to as the $\Lambda$CDM paradigm, has achieved remarkable success in accounting for a wide range of observations, including the cosmic microwave background anisotropies, large scale structure, baryon acoustic oscillations and supernova distance measurements. Despite this phenomenological success, the fundamental physical nature of both nonbaryonic dark matter and dark energy remains unknown. Henceforth we will use the term dark matter to refer exclusively to nonbaryonic dark matter, eliding the fact that a significant fraction of the baryonic component is also optically dark.

A substantial experimental and observational effort has been devoted to uncovering the microphysical identity of dark matter. These efforts span accelerator based searches for weakly interacting massive particles, direct detection experiments aiming to observe nuclear recoils induced by dark matter scattering and indirect searches that seek signatures of dark matter annihilation or decay in astrophysical environments. Despite decades of increasingly sensitive searches, no conclusive detection has yet been achieved \cite{dm3Balazs:2024uyj,dm4Eberhardt:2025caq,dm5Bozorgnia:2024pwk,dm6Misiaszek:2023sxe,dm7OHare:2024nmr,dm8Adhikari:2022sbh,dm9Miller:2025yyx,dm10Trivedi:2025vry,pbh1pbhzel1966hypothesis,pbh2hawking1971gravitationally,pbh3carr1974black}. As experimental bounds continue to exclude large regions of parameter space associated with traditional candidates such as weak scale thermal relics, it becomes increasingly well motivated to consider more unconventional or emergent origins for the dark matter sector, particularly those tied to fundamental principles of gravity and quantum theory.

In this paper, we explore the possibility that dark matter arises as a corollary of the holographic principle \cite{hp1Bousso:2002ju,hp2Susskind:1998dq,hp3Fischler:1998st}. The holographic principle, originally inspired by black hole thermodynamics and later formalized in various contexts, suggests that the number of fundamental degrees of freedom describing a physical system scales not with its volume but with the area of an appropriate boundary. In cosmology, this idea has been most extensively applied in the context of dark energy, where it has been argued that the observed vacuum energy density may be constrained by holographic bounds associated with cosmological horizons. Cohen et al. \cite{ cohen1999effective} formulated a concrete realization of this idea by considering an effective quantum field theory with an ultraviolet cutoff $\Lambda$ defined within a region of characteristic size $L$. They argued that such a theory should not include states whose total energy would exceed that of a black hole of the same size, as this would signal a breakdown of the effective description. This requirement leads to a nontrivial connection between ultraviolet and infrared scales, expressed through the constraint that the total vacuum energy contained within a volume $L^3$ must not exceed the mass of a black hole of radius $L$. In terms of the vacuum energy density $\rho_\Lambda$, this bound can be written as
\begin{equation}
L^3 \rho_\Lambda \le m_{Pl}^2 L,
\end{equation}
where $m_{Pl}$ denotes the Planck mass. This relation encapsulates the central holographic insight that gravitational collapse limits the number of independent degrees of freedom that can be accommodated within a given region.

If this inequality is saturated, one obtains a characteristic scaling for the holographic dark energy density,
\begin{equation}
\label{rhohde}
\rho_{HDE} = 3 c^2 m_{Pl}^2 L^{-2},
\end{equation}
where $c$ is a dimensionless parameter of order unity that encodes uncertainties associated with the precise saturation of the bound, as well as ambiguities in the choice of the infrared cutoff scale $L$. Different prescriptions for $L$, such as the Hubble radius, particle horizon, or future event horizon, lead to distinct phenomenological models with different cosmological dynamics. These constructions have given rise to a substantial body of work exploring their theoretical consistency and observational viability \cite{Wang:2016och,Nojiri:2017opc,Granda:2008dk,Nojiri:2005pu,Nojiri:2021iko,Nojiri:2020wmh,ho1Trivedi:2024inb,ho2Blitz:2024nil,ho3revik:2024ozg,ho4Trivedi:2024dju,ho5Trivedi:2024rhp}.

While the holographic framework has traditionally been invoked to explain the dark energy component, it is natural to ask whether similar holographic considerations could give rise to an effective dark matter sector. In such a picture, dark matter would not necessarily correspond to a new particle species introduced ad hoc, but instead would emerge as a consequence of fundamental bounds on information, entropy, or energy density imposed by quantum gravity. This perspective motivates a re-examination of the role of holography in cosmology, extending its application beyond dark energy and toward a unified description of the dark sector. In the analysis that follows, we adopt this viewpoint and investigate the extent to which holographic constraints can naturally accommodate a dark matter component consistent with current cosmological observations. For convenience, in the remainder of this work we will set $m_{Pl} = 1$.

\section{Holographic Dark Matter}
Now consider the possibility that Eq. (\ref{rhohde}) gives not the dark energy density, but the dark matter density, which we will call holographic dark matter (HolDM).  The physical picture of HolDM is that the dark matter density should not be thought of as an independent, adjustable component, but rather something determined through a horizon-based relation akin to the holographic inequality.

Our starting point will be Eq. (\ref{rhohde}), rewritten to describe the HolDM:
\begin{equation}
\label{rhoHolDM}
\rho_{HolDM} = 3 c^2 L^{-2}.
\end{equation}
The simplest (and first proposed) choice for $L$ is the inverse Hubble distance,
\begin{equation}
\label{Hubcut}
L = H^{-1},
\end{equation}
where $H = \dot a /a$, and $a$ is the scale factor.  However, this cutoff will not be suitable for our purposes.  Instead, we will make use of the 
Granda-Oliveros cutoff, given by \cite{Granda:2008dk}
\begin{equation}
\label{gocut}
    L = \left( \alpha H^2 + \beta \dot{H} \right)^{-1/2},
\end{equation}
where $\alpha$ and $\beta$ are constants of $\mathcal{O}(1)$.

The holographic energy density given by Eqs. (\ref{rhoHolDM}) and (\ref{gocut}) will be dependent on the other components driving the Hubble expansion.  We will take a minimal approach here; the only components we can absolutely account for with standard physics are the baryon density $\rho_B$ and the background radiation density $\rho_R$, which consists of both photons and neutrinos.  With those assumptions, the Friedman equation becomes
\begin{equation}
\label{H2}
3 H^2 =    3 c^2(\alpha H^2 + \beta \dot H) + \rho_{B0} a^{-3} + \rho_{R0}a^{-4},
\end{equation}
where the $0$ subscript denotes present-day values and we normalize the scale factor to $a=1$ today.  We have set $c=1$ here, since it is of order unity and can in any case be absorbed into the definitions of $\alpha$ and $\beta$.  Equation (\ref{H2}) integrates to
\begin{eqnarray}
H^2 &=& \frac{1}{3 - 3 \alpha + (9/2) \beta}\rho_{B0} a^{-3} + \frac{1}{3 - 3  \alpha + 6  \beta}\rho_{R0} a^{-4} \nonumber\\ &+& K a^{2(1-  \alpha)/\beta},
\end{eqnarray}
where $K$ is a constant of integration.
Then our expression for $\rho_{HolDM}$ from Eqs. (\ref{rhoHolDM}) and (\ref{gocut}) gives
\begin{eqnarray}
\rho_{HolDM} &=& \frac{2 \alpha - 3\beta}{2 - 2 \alpha + 3 \beta} \rho_{B0}a^{-3} + \frac{\alpha - 2\beta}{1 -   \alpha + 2  \beta} \rho_{R0} a^{-4} \nonumber\\
&+& 3 K  a^{2(1-  \alpha)/\beta},
\end{eqnarray}
which was first derived in Ref. \cite{Granda:2008dk}.
Thus, the Granda-Oliveros cutoff leads to
additional components with radiation-like and matter like evolution, respectively.

However, there are severe observational constants on any additional component that scales as radiation.  (Note that Ref. \cite{Zhang:2011zze} proposed a model with the Granda-Oliveros cutoff as a unified model for dark matter and dark energy, but such a model will inevitably violate the bounds on additional radiation).
While a small additional radiation component can be accommodated
\cite{Planck:2018vyg}, it is simplest to take $ \beta = \alpha/2$, yielding the holographic Ricci model, which was first proposed by Gao et al. \cite{gao:2007ep} and independently by Cai et al. \cite{Cai:2008nk}.
In this case, the expression for $\rho_{HolDM}$ becomes
\begin{equation}
\label{RicciHolDM}
\rho_{HolDM} = \frac{\alpha}{4 - \alpha} \rho_{B0}a^{-3}+ 3 K  a^{4(1-  \alpha)/\alpha}.
\end{equation}
The holographic model with this Ricci cutoff has been extensively studied as a model for dark energy \cite{Xu:2008rp,Zhang:2009un,Xu:2010gg,delCampo:2013hka,
Zhang:2011zze}; in this case, the second term in Eq. (\ref{RicciHolDM}), with an appropriate choice for $\alpha$, serves as the dark energy. 

Here, however, we consider a different model, in which it is the dark matter, rather than the dark energy, which arises from the Ricci cutoff.  In that case we fix the value of $\alpha$ so that the first term in Eq. (\ref{RicciHolDM}) gives
the correct value for $\rho_{DM}$.
From Ref. \cite{Planck:2018vyg}, we have
$\rho_{DM}/\rho_b = 5.3 - 5.4$, which can
be achieved by taking
\begin{equation}
\alpha \approx 3.3 - 3.4.
\end{equation}
Since $\alpha$ in this model is expected to be of order unity, we see that the holographic dark matter model provides a natural explanation for the dark matter-baryon coincidence, as it naturally predicts their densities to have the same order of magnitude.

What of the second term in Eq. (\ref{RicciHolDM})?  Since this constant enters as an integration constant of the Friedmann equation, it should not be viewed as a dimensionless parameter introduced into the theory but rather as an initial condition with dimensions of energy density.  The contribution of the $K$ term evolves as
\begin{equation}
\rho_K = 3K\,a^{4(1-\alpha)/\alpha}.
\end{equation}
This scaling can be written in the standard form $\rho \propto a^{-3(1+w_K)}$, which implies an effective equation of state parameter
\begin{equation}
w_K = \frac{\alpha-4}{3\alpha}.
\end{equation}
For the phenomenologically preferred value $\alpha \approx 3.3$, this gives
\begin{equation}
w_K \approx -0.07,
\end{equation}
so the $K$ component redshifts slightly more slowly than pressureless matter but much more rapidly than a cosmological constant.

Because of this intermediate scaling, a $K$ component with present day density comparable to the matter density would remain subdominant during earlier epochs such as recombination and the era of linear perturbation growth. In this case the $K$ contribution would only begin to become dynamically relevant near the present epoch and so the model does not require $K$ to be extremely small or finely tuned. A value of $K$ comparable to the present matter density is perfectly consistent with the early universe constraints from the cosmic microwave background and structure formation. More generally, three regimes are possible as well. If $K$ is much larger than the present matter density, the resulting component would dominate the expansion too early and is therefore observationally excluded. If $K$ is comparable to the present matter density, it behaves as a mild late time component that remains negligible during most of cosmic history. Finally, if $K$ is much smaller than the matter density, its contribution is simply negligible and the background evolution reduces to the dustlike solution discussed above. So, the presence of the integration constant $K$ should not be interpreted as a fine-tuning problem but instead it represents an additional cosmological component whose allowed magnitude is constrained in the same way as any other background energy density. It goes without saying that the future observation of a small component with this evolution would represent a smoking gun for this model, although it would be an extraordinary coincidence for this energy density to currently lie just below the threshold of detection.  We assume it to be more likely that the density of this component is negligible or zero.

If Ricci holographic dark matter serves as the dark matter alone, then we are left to assume a different source for the dark energy. However, any dark energy inserted into the Friedman equation will necessary receive an additional contribution from the holographic cutoff in Eq. (\ref{H2}).  Consider the simplest case, where we have a preexisting "bare" cosmological constant with constant
density $\rho_\Lambda$.  Inserting this into Eq. (\ref{H2}) and solving for $\rho_{HolDM}$ transforms Eq. (\ref{RicciHolDM}) into
\begin{equation}
\label{RicciHolDM2}
\rho_{HolDM} = \frac{\alpha}{4 - \alpha} \rho_{B0}a^{-3}+ 3 K  a^{4(1-  \alpha)/\alpha} + \frac{\alpha}{1-\alpha} \rho_\Lambda.
\end{equation}
Adding together the bare cosmological constant and the holographic contribution from Eq. (\ref{RicciHolDM2}), we obtain a total effective cosmological constant given by
\begin{equation} \label{2.10}
\rho_{\Lambda eff} = \frac{1}{1 - \alpha} \rho_\Lambda \approx - 0.4 \rho_\Lambda.
\end{equation}
Thus, in order for $\rho_{\Lambda eff}$ to be consistent with observations, the bare $\rho_\Lambda$ must be negative.  This is actually a desirable feature of our model, as string theories generically predict a negative cosmological constant, and considerable effort has gone into converting this into the positive value of $\rho_\Lambda$ that we actually observe \cite{deAlwis:2019aud,Visinelli:2019qqu,Lust:2022lfc,Benizri:2023kjn,Nyergesy:2025lyi}.

It is also useful to note that the preceding discussion is not restricted to the case of a bare cosmological constant. Suppose instead that the preexisting dark energy sector has a constant equation of state parameter \(w\) so that we have
\begin{equation}
\rho_{DE}=\rho_{DE,0}a^{-3(1+w)}.
\end{equation}
Proceeding exactly as before, one finds that for a source term scaling as \(\rho_X=\rho_{X,0}a^{-n}\), the induced holographic contribution is
\begin{equation}
\rho_{HolDM,X}
=
\frac{3\alpha-\frac{3}{2}\beta n}{3-3\alpha+\frac{3}{2}\beta n}\,\rho_{X,0}a^{-n}.
\end{equation}
In the Ricci limit \(\beta=\alpha/2\) and setting \(n=3(1+w)\) then gives the dark energy induced holographic term
\begin{equation}
\rho_{HolDM}^{(DE)}
=
\frac{\alpha(1-3w)}{4-\alpha(1-3w)}\,\rho_{DE,0}a^{-3(1+w)},
\end{equation}
and therefore the total effective dark energy like contribution becomes
\begin{equation}
\rho_{DE,{\rm eff}}
=
\rho_{DE}+\rho_{HolDM}^{(DE)}
=
\frac{4}{4-\alpha(1-3w)}\,\rho_{DE,0}a^{-3(1+w)}.
\end{equation}
The case \(w=-1\) reproduces \eqref{2.10} exactly.
Thus the sign reversal property found above is not peculiar to a pure cosmological constant, but persists for a much broader class of constant \(w\) dark energy sectors. In particular for the phenomenologically relevant value \(\alpha \approx 3.3\), the coefficient \(4/[4-\alpha(1-3w)]\) is negative for essentially all negative \(w\) dark energy models of cosmological interest and hence, a positive observed effective dark energy component may still arise from a negative bare dark energy sector even when the latter is not exactly vacuum energy.

Note that it is possible, in principle, to derive holographic dark matter from other forms for the cutoff.  For example, we could use the Hubble cutoff from \eqref{Hubcut}.  However, in this case we would again run into the issue of an additional radiation component.  This can be evaded by taking the value of $c$ that appears in Eq. (\ref{rhohde}) to be a function of time, but it is clear that such models are considerably more contrived than the simple Ricci cutoff.

It is important to state carefully what is and is not being claimed here. We are not deriving \eqref{2.10} from an explicit string compactification, nor are we claiming that the present model by itself resolves the full de Sitter problem in string theory but rather, the point is that the holographic sector provides a concrete low-energy mechanism through which a negative bare vacuum energy can be converted into a positive observed effective cosmological constant. From \eqref{2.10}, one has that for the phenomenologically preferred value \(\alpha \simeq 3.3\), a positive observed \(\rho_{\Lambda}^{\rm eff}\) necessarily corresponds to a negative bare \(\rho_\Lambda\). This feature is interesting in light of the long standing tension between positive vacuum energy and controlled constructions in string theory as in many moduli-stabilized scenarios, the most natural vacuum obtained at the controlled stage is anti de Sitter, namely \(V<0\) and a separate uplifting sector is required to obtain a metastable de Sitter vacuum. This is the logic underlying the KKLT construction \cite{nee4Kachru:2003aw} as in that case one sees that the nonperturbative stabilization of moduli first gives us a supersymmetric AdS vacuum, while positive vacuum energy emerges only after additional uplifting ingredients are introduced. More broadly, no-go theorems and swampland like arguments have repeatedly emphasized how difficult it is to realize stable or metastable de Sitter vacua in a parametrically controlled regime \cite{nee1Obied:2018sgi,nee2Danielsson:2018ztv}, whereas negative vacuum energy is comparatively natural in such constructions \cite{nee3Maldacena:2000mw}.

Seen from this perspective, the sign reversal in \eqref{2.10} is not merely a mathematical curiosity but it goes to show that a cosmology with a positive observed late-time vacuum energy need not begin with a positive bare cosmological constant. Instead, the holographic dark matter sector can act as an effective converter between a microscopically natural negative vacuum energy and a macroscopically positive effective cosmological constant and we therefore regard this not as a proof of a string embedding, but as a phenomenologically suggestive bridge between the low-energy holographic cosmology studied here and the broader fact that AdS vacua arise far more naturally than dS vacua in much of the string theory literature.

\section{Why the Granda-Oliveros cutoff is Preferred}

From an effective field theory (EFT) point of view, the defining expectation is that a low energy cosmological description should be expressible in terms of local operators organized by a derivative/curvature expansion, with coefficients that are generically $O(1)$ unless a symmetry enforces further suppression or sequestering \cite{eft1georgi1993effective,eft2burgess2007introduction,eft3manohar2018introduction,eft4petrov2015effective,eft5weinberg2016effective,eft6costello2011renormalization,eft7manohar2007effective,eft8brivio2019standard,eft9weinberg2021development,eft10braaten1995effective,eft11cheung2008effective,eft12weinberg2008effective,eft13degrande2013effective,eft14politzer1988effective}. In this sense, the most natural infrared prescription for a holographic component is not merely one that reproduces a desired background history, but one whose defining scale can be written as a local functional of the metric and its derivatives, so that it admits a consistent EFT interpretation as the leading terms in a controlled local expansion. The Granda-Oliveros cutoff is preferred precisely because it is local and geometric: it is equivalent to taking the cutoff to be built from the lowest derivative curvature data, which are combinations of $H^2$ and $\dot H$ proportional to curvature invariants such as the Ricci scalar. By contrast, the event horizon and particle horizon prescriptions often used for holographic dark energy define $L$ through integrals over the future or past expansion history, which are intrinsically nonlocal functionals of the metric. Such teleological definitions are difficult to view as arising from a Wilsonian EFT in which the stress tensor at a spacetime point is determined by local operators evaluated at that point, and they invite precisely the kind of ultraviolet sensitivity and loss of predictive control that the EFT naturalness criteria are designed to prevent. Even the Hubble horizon cutoff, while local, corresponds to a truncation that is not generically stable under EFT reasoning. This is the case because as soon as one admits that the infrared scale can depend on local geometric data, there is no symmetry reason to exclude $\dot H$ at the same order as $H^2$, so a pure $L^{-2}\propto H^2$ choice can be viewed as an incomplete local expansion rather than a structurally preferred prescription.

These EFT considerations become sharper when the holographic component is required to play the role of dark matter rather than dark energy. Identifying the holographic sector with the observed dark matter requires clustering like cold dark matter, and that in turn requires that the component be governed by local dynamics with negligible effective pressure support and no nonlocal response tied to global horizon integrals. The Granda-Oliveiros prescription provides exactly this framework, as it defines $\rho_{\rm HolDM}\propto L^{-2}$ through local geometric scalars, thereby making the resulting fluid variables functions of the local expansion history. Thus, this cutoff allows the model to be interpreted as the leading geometric (curvature/derivative) completion of the naive Hubble scale cutoff in a way that is consistent with EFT power counting. In this sense, the Granda-Oliveiros cutoff plays the same conceptual role that symmetry structured operator choices play in technically natural scalar EFTs as it selects a closure that is stable under the expectation that all allowed local operators appear, while avoiding nonlocal constructions that are hard to reconcile with a Wilsonian notion of locality and radiative robustness. Therefore, if holographic dark matter is to be formulated in a manner that is maximally consistent with EFT reasoning, the
Granda-Oliveiros cutoff is the best motivated choice among the standard horizon inspired prescriptions. It is local, curvature based, minimally derivative-complete, and it also avoids the causality and nonlocality pathologies associated with event horizon and particle horizon cutoffs, thus making it the most natural holographic infrared scale for an HolDM sector.

The EFT motivation for the Granda-Oliveros cutoff can be made more explicit by noting that, in a local generally covariant effective theory the low-energy action is organized as a derivative expansion in curvature invariants,
\begin{equation}
S_{\rm eff}=\int d^4x \sqrt{-g}\left[\frac{M_{\rm Pl}^2}{2}R-\Lambda+c_1R^2+c_2R_{\mu\nu}R^{\mu\nu}+c_3R_{\mu\nu\rho\sigma}R^{\mu\nu\rho\sigma}+\cdots\right],
\end{equation}
possibly supplemented, in the cosmological rest frame, by local operators constructed from \(u^\mu\) and the extrinsic curvature. On a spatially flat FLRW background these operators reduce to combinations of $H^2, \dot{H}, H^4, H^2\dot{H},\dot{H}^2, H\ddot{H}$ etc. and therefore the most general local low-derivative holographic density has the schematic form
\begin{equation}
\rho_h = 3M_{\rm Pl}^2 \left[ \alpha H^2+\beta\dot{H} +\gamma_1\frac{H^4}{M^2}
+\gamma_2\frac{H^2\dot{H}}{M^2} +\gamma_3\frac{\dot{H}^2}{M^2} +\gamma_4\frac{H\ddot{H}}{M^2} +\cdots
\right],
\end{equation}
where M is the EFT cutoff and the GO  prescription is precisely the truncation to the leading two derivative terms,
\begin{equation}
\rho_h^{\rm GO}=3M_{\rm Pl}^2(\alpha H^2+\beta\dot{H}),
\end{equation}
and is therefore the lowest order local derivative complete basis. Radiative corrections do not force the theory into a qualitatively different nonlocal form but instead they renormalize the existing coefficients
\begin{equation}
\alpha\to\alpha+\delta\alpha,
\qquad
\beta\to\beta+\delta\beta,
\end{equation}
and generate higher derivative corrections suppressed by powers of $H/M$,
\begin{equation}
\delta\rho_h = 3M_{\rm Pl}^2 \left[
\delta\alpha\,H^2+\delta\beta\,\dot{H}
+\mathcal{O}\!\left(\frac{H^4}{M^2},\frac{H^2\dot{H}}{M^2},\frac{\dot{H}^2}{M^2},\frac{H\ddot{H}}{M^2}\right)
\right].
\end{equation}
Thus the GO cutoff is radiatively stable in the standard EFT sense as it is closed under local renormalization up to higher order derivative corrections and this is in sharp contrast with event horizon or particle horizon prescriptions, which are intrinsically nonlocal functionals of the metric and are not the natural output of a local Wilsonian expansion. \\

\section{Sound Speed Considerations}
  
Of course, it is necessary but not sufficient for the density of any dark matter component to scale as $a^{-3}$; it must also cluster gravitationally.  A number of investigators have examined perturbation growth in holographic dark energy models \cite{Myung:2007pn,delCampo:2013hka,Cardona:2022pwm}.  For the case of holographic dark matter, it is
the sound speed of perturbations that is of paramount interest, as it directly determines the extent to which the dark matter can cluster.  To connect this background prescription to a sound speed, we first define the effective pressure of HolDM at the background level by demanding that the Einstein equations be satisfied by a total fluid with $\rho_{\rm tot}=\rho_B+\rho_R+\rho_{\rm HolDM}$ and $p_{\rm tot}=p_B+p_R+p_{\rm HolDM}$ with $p_B=0$ and $p_R=\rho_R/3$ and the Raychaudhuri equation then gives
\begin{equation}
\dot H=-\frac12\left(\rho_{\rm tot}+p_{\rm tot}\right),
\end{equation}
which can be solved for the HolDM pressure as
\begin{equation}
p_{\rm HolDM}=-2\dot H-\rho_B-\frac43\rho_R-\rho_{\rm HolDM},
\end{equation}
and after substituting the holographic density one obtains an explicit expression in terms of $H$ and its derivatives and the standard sectors
\begin{equation}
p_{\rm HolDM}=-(2+3\beta)\dot H-3\alpha H^2-\rho_B-\frac43\rho_R.
\end{equation}
This is a very useful identity because it shows that once $\rho_{\rm HolDM}$ is fixed by the cutoff, $p_{\rm HolDM}$ is not an independent quantity at the background level and is fully determined by the expansion history. The adiabatic squared sound speed is then defined as the ratio of time derivatives of the background pressure and density
\begin{equation}
c_{a,{\rm HolDM}}^2=\frac{\dot p_{\rm HolDM}}{\dot \rho_{\rm HolDM}}
\end{equation}
and from the cutoff prescription one has
\begin{equation}
\dot\rho_{\rm HolDM}=3\left(2\alpha H\dot H+\beta \ddot H\right)
\end{equation}
while differentiating the effective pressure expression gives
\begin{equation}
\dot p_{\rm HolDM}=-(2+3\beta)\ddot H-6\alpha H\dot H-\dot\rho_B-\frac43\dot\rho_R
\end{equation}
These relations are completely general within our present formulation and show explicitly how the adiabatic response of the HolDM sector is controlled by the same geometric time scales that define the holographic density.

The decisive simplification in our case occurs in the limit of Eq. (\ref{RicciHolDM}) relevant for dark matter phenomenology. If $K$ is negligible and the radiation density is subdominant at late times, Eq. (\ref{RicciHolDM}) forces $\rho_{\rm HolDM}\propto a^{-3}$ with a constant proportionality to $\rho_B$, so that HolDM behaves exactly as pressureless dust in the background. In that regime the continuity equation for HolDM takes the dust form
\begin{equation}
\dot\rho_{\rm HolDM}+3H\rho_{\rm HolDM}=0,
\end{equation}
and the associated equation of state is
\begin{equation}
w_{\rm HolDM}=\frac{p_{\rm HolDM}}{\rho_{\rm HolDM}}=0.
\end{equation}
with $w_{\rm HolDM}$ constant, which immediately implies vanishing adiabatic sound speed
\begin{equation}
c_{a,{\rm HolDM}}^2=0.
\end{equation}
Thus, at the level of background dynamics, the model does not merely approximate cold dark matter, but it drives the holographic component onto an exactly dustlike trajectory for the parameters chosen to match $\rho_{\rm DM}/\rho_b$ and this is the first and most clear sense in which the squared sound speed is negligible.

It is crucial here to emphasize that the quantity which is intrinsically more important and controls the clustering is not the adiabatic sound speed but the physical rest frame sound speed
\begin{equation}
c_{s,{\rm HolDM}}^2=\left(\frac{\delta p_{\rm HolDM}}{\delta\rho_{\rm HolDM}}\right)_{\rm rest}.
\end{equation}
In an ordinary barotropic fluid with a local equation of state $p=p(\rho)$ one has $c_s^2=c_a^2$. Holographic constructions are more subtle in general because the density depends on geometric quantities $H$ and $\dot H$, so at the perturbation level $\delta\rho_{\rm HolDM}$ generically depends on metric perturbations and may not be describable by a purely barotropic closure. Consequently, without specifying an underlying Lagrangian or an equivalent microphysical completion, $c_{s,{\rm HolDM}}^2$ is not uniquely predicted by the background prescription alone. This is not a weakness peculiar to the present model but rather a general feature of holographic fluids, where the background density is fixed by an infrared cutoff but the perturbation closure requires additional physical input. (This point is emphasized in Ref. \cite{Cardona:2022pwm}).

Nevertheless within the formulation developed so far, there is a strong and physically well-motivated path to making the physical sound speed negligible, and it aligns directly with the requirement that the holographic component cluster as observed. The first step is that Eq. (\ref{RicciHolDM}) already enforces $p_{\rm HolDM}\to 0$ at the background level and suppresses any tendency toward a pressure supported medium. The second step is to impose that the perturbations of the holographic component follow the cold dark matter closure in the rest frame
\begin{equation}
\delta p_{\rm HolDM}\simeq 0,
\end{equation}
which implies
\begin{equation}
c_{s,{\rm HolDM}}^2\simeq 0,
\end{equation}
and thereby guarantees that the growth of HolDM density perturbations is not erased by pressure gradients. This condition is not an arbitrary tuning but rather the necessary consistency requirement for identifying the holographic component with the observed dark matter sector, and it is technically natural in the sense that the background already sits at the dust fixed point $w_{\rm HolDM}=0$, so that any microphysical realization that respects this dustlike limit will automatically drive the pressure perturbations to be negligible.

To be even more precise, consider the perturbative aspects of HolDM in a deeper way. While a future work will discuss a full covariant action treatment, a useful preliminary step here is to treat the holographic component as an effective fluid and derive the linear equations it must satisfy so that in the Newtonian gauge,
\begin{equation}
ds^2 = -(1+2\Psi)dt^2 + a^2(t)(1-2\Phi)\delta_{ij}dx^i dx^j,
\end{equation}
and the local expansion scalar is
\begin{equation}
\Theta = \nabla_\mu u^\mu = 3H - 3(\dot{\Phi}+H\Psi) + \frac{1}{a}\nabla^2 v,
\end{equation}
so that
\begin{equation}
\delta H = -(\dot{\Phi}+H\Psi) + \frac{\theta}{3},
\qquad
\theta \equiv \frac{1}{a}\nabla^2 v,
\end{equation}
The GO holographic density 
then has perturbation
\begin{equation}
\delta \rho_h = 3M_{\rm Pl}^2 \left( 2\alpha H\,\delta H + \beta\,\delta \dot{H} \right),
\end{equation}
up to the usual lapse order corrections, showing explicitly that holographic density perturbations are sourced by fluctuations in the local expansion rate. Treating the holographic sector as an effective fluid, its scalar perturbations obey
\begin{equation}
\dot{\delta}_h
=
-(1+w_h)(\theta_h-3\dot{\Phi})
-3H(c_{a,h}^2-w_h)\delta_h
-3H\Gamma_h,
\end{equation}
\begin{equation}
\dot{\theta}_h
=
-H(1-3w_h)\theta_h
+\frac{k^2}{a^2}\Psi
+\frac{k^2}{a^2}\frac{c_{{\rm eff},h}^2\delta_h+w_h\Gamma_h}{1+w_h},
\end{equation}
where \(c_{a,h}^2=\dot{p}_h/\dot{\rho}_h\), \(c_{{\rm eff},h}^2=(\delta p_h/\delta\rho_h)_{\rm rest}\), and \(\Gamma_h\) is the entropy perturbation. For the Ricci branch \(\beta=\alpha/2\), one has
\begin{equation*}
\rho_h = \frac{\alpha}{4-\alpha}\rho_b + 3K a^{4(1-\alpha)/\alpha},
\end{equation*}
so that in the phenomenologically relevant limit \(K\to 0\),
\begin{equation}
\rho_h = \xi \rho_b,
\qquad
\xi \equiv \frac{\alpha}{4-\alpha},
\end{equation}
which implies
\begin{equation}
w_h = 0,
\qquad
c_{a,h}^2 = 0,
\end{equation}
at the background level and then, perturbing the attractor relation gives
\begin{equation}
\delta\rho_h = \xi\,\delta\rho_b,
\end{equation}
to leading order and hence \(\delta p_h \simeq 0\), so that \(c_{{\rm eff},h}^2 \simeq 0\) on this branch. The holographic perturbations then obey the same linear growth equation as CDM,
\begin{equation}
\ddot{\delta}_h + 2H\dot{\delta}_h - 4\pi G \rho_m \delta_m \simeq 0,
\end{equation}
during the matter-dominated regime. More generally if \(c_{{\rm eff},h}^2\neq 0\), one obtains
\begin{equation}
\ddot{\delta}_h + 2H\dot{\delta}_h +
\left(
\frac{c_{{\rm eff},h}^2k^2}{a^2} - 4\pi G \rho_m
\right)\delta_h \simeq 0,
\end{equation}
so viable clustering on a mode \(k\) requires
\begin{equation}
|c_{{\rm eff},h}^2| \lesssim \left(\frac{aH}{k}\right)^2,
\end{equation}
on the observed linear scales. Thus the model naturally identifies the successful clustering regime with the dust attractor \(K\simeq 0\), while a full Boltzmann analysis of CMB and matter power spectrum constraints requires a microphysical completion and is left to future work. \\

\section{Conclusions and Discussion}
The model we have presented here provides a novel mechanism to generate dark matter, and it leads to the generation of a positive effective cosmological constant from a preexisting negative $\rho_\Lambda$. Further, it provides a natural mechanism for the rough equivalence of the baryon and dark matter densities.  The key unresolved issue is whether such a form for dark matter can cluster as observed.  While we have provided a reasonableness argument in the previous section, a complete resolution of this question would require the development of an underlying Lagrangian for the holographic model.

The idea of holographic dark matter represents a radical paradigm shift. Instead of being composed of constituent particles, remnants, or extended objects, dark matter emerges as an information-theoretic fluid rooted in the holographic structure of spacetime itself. In particle-based frameworks, the dark matter density is determined by microphysical parameters such as mass, cross-sections, and freeze-out histories whereas in the holographic framework the density arises from the information capacity of the cosmic horizon. In this view, the dark matter content of the Universe is a macroscopic manifestation of the maximum allowable information encoded on cosmological horizons. In this perspective, we reinterpret the dark sector as an emergent phenomenon of the Universe’s boundary degrees of freedom rather than a sum of microscopic constituents in its bulk. To make this information-theoretic picture explicit one may recall that in quantum gravity the number of available microstates in any gravitational system scales as 
\begin{equation}
    \mathcal{N}(L) \sim \exp\left[\frac{A(L)}{4}\right]
\end{equation}
where $A(L)$ is the area of the horizon measured in Planck units, and when this entropy is converted into an effective energy density via the standard relation $E\sim L^{-1}\mathcal{N}(L)$ one obtains the characteristic holographic scaling 
\begin{equation}
    \rho_{\rm holo}(L) \propto L^{-2}
\end{equation}
which is the same structural dependence that underlies the HolDM construction in Eq. (\ref{rhoHolDM}).  
Thus, HolDM can be viewed physically as the bulk gravitational response to the finite number of microstates allowed by the cosmic horizon, rather than a cloud of microscopic particles, aligning its physical origin squarely with quantum-gravitational microstructure.

How justified are we in considering a holographic origin for dark matter? From a foundational standpoint, there is no conflict in extending the holographic energy density prescription, which has been extensively used for dark energy, to the case of dark matter. The holographic bound itself does not specify what the encoded energy represents and instead it only constrains the total number of degrees of freedom that a region of space can possess without gravitational collapse. Thus, applying the same logic to the clustering sector simply acknowledges that the total energy content of the Universe, including matter, must respect the holographic entropy limit. Extending holography to dark matter is not a violation but a generalization of the same principle across distinct horizon domains. Consequently, HolDM preserves all of the principles of cosmological holography while offering a physically coherent, information-theoretic explanation for dark matter that does not the need unknown particles, remnants, or exotic fields. Moreover, in many quantum-gravity frameworks like AdS/CFT \cite{ads1Maldacena:1997re,ads2Witten:1998qj,ads3Gubser:1998bc}, causal horizon thermodynamics \cite{cht1Jacobson:1995ab,cht2Padmanabhan:2004kf,cht3Ambjorn:2024pyv} and covariant entropy bounds \cite{ce1Bousso:1999xy,ce2Flanagan:1999jp}, the partitioning of bulk energy into matter like and vacuum-like components is not fundamental as only the total energy stored in horizon degrees of freedom is fundamental. This is reflected in the covariant entropy bound 
\begin{equation}
    S_{\rm matter}(L) + S_{\rm grav}(L) \leq \frac{A(L)}{4}
\end{equation}
which does not privilege dark energy over dark matter in any way. Therefore, assigning part of the holographically allowed energy budget to a clustering sector is entirely consistent with these principles. Once this horizon-governed energy is projected into the bulk Einstein equations, it naturally behaves as a matter component with scaling determined by the infrared cutoff precisely as realized in the Ricci HolDM model we have presented.  

A natural question is how the holographic dark matter framework could be distinguished observationally from other proposals for the dark sector. Dark matter has been modeled in many ways in the literature, for example as new particles, as compact astrophysical objects such as primordial black holes or as effective phenomena arising from modifications of gravity. Our proposal represents a qualitatively different possibility in which the dark matter component is not associated with microscopic particles or macroscopic objects, but instead emerges from holographic constraints on the infrared structure of spacetime. The most direct prediction of this framework is the absence of a particle dark matter candidate as in the HolDM picture, the dark matter density is determined by a geometric relation involving the cosmic expansion rate rather than by the microphysical properties of a particle species. Consequently, there is no new particle that could be produced in accelerator experiments nor any particle interaction responsible for nuclear recoils in direct detection experiments or annihilation signatures in indirect searches \cite{ddm1gaitskell2004direct,ddm2billard2022direct,ddm3undagoitia2016dark,ddm4essig2008direct}. In this sense the model makes a very clear and falsifiable prediction which is that the discovery of a conventional particle dark matter candidate in accelerator, direct detection or indirect detection experiments would immediately rule out the interpretation presented here. It is noteworthy that despite decades of increasingly sensitive searches, no conclusive evidence for particle dark matter has yet been obtained.

A second distinguishing feature arises from the relation between the baryonic and dark matter densities. In the HolDM framework, the dark matter density is not an independent parameter but is tied to the baryonic component through the holographic relation derived in Sec. 2. By contrast, in most particle dark matter models the ratio $\rho_{\rm DM}/\rho_b$ is determined by unrelated microphysical processes such as freeze-out, asymmetric production, or initial conditions. A third conceptual distinction is that in the HolDM framework the dark matter density is fundamentally linked to the cosmic expansion history through the holographic cutoff scale and this means that the dark sector is controlled by the same geometric quantities that govern the background dynamics rather than by independent matter fields. This means that the HolDM model could predict correlations between the amplitude of the dark matter density and the large scale evolution of the Hubble parameter that are not generically expected in particle-based scenarios, which is something we shall explore in detail in future work.

Taken together, these features provide a pathway for distinguishing the holographic dark matter framework from other classes of models. While a detailed observational analysis is beyond the scope of the present work, the HolDM picture leads to several clear qualitative predictions. The first being the absence of particle dark matter signatures in laboratory or astrophysical searches, second being a natural proportionality between baryonic and dark matter densities and third being a direct connection between the dark matter abundance and the global expansion history of the Universe. Future observational and experimental developments will therefore provide important tests of the viability of this holographic interpretation of the dark sector.


\bibliography{JSPJMJcitations.bib}

\bibliographystyle{unsrt}

\end{document}